\documentclass[twocolumn,prl,superscriptaddress,showpacs]{revtex4}
\usepackage{graphicx,subfigure,amsmath,amssymb,hyperref,mathrsfs}
\usepackage{setspace}

\usepackage{color}

\newcommand{\ket}[1]{|{#1}\rangle}
\newcommand{\bn}{\begin{eqnarray}}
\newcommand{\en}{\end{eqnarray}}
\newcommand{\n}{\nonumber}
\begin{document}

\title{Boosting the Rotational Sensitivity of Matter-wave Interferometry with Nonlinearity}

\author{M. Kol\'a\v r}
\affiliation{Department of Optics, Palack\'{y}
University, 771 46 Olomouc, Czech Republic}
\author{T. Opatrn\'y}
\affiliation{Department of Optics, Palack\'{y}
University, 771 46 Olomouc, Czech Republic}
\author{Kunal K. Das}
\affiliation{Department of Physical Sciences, Kutztown University of Pennsylvania, Kutztown, Pennsylvania 19530, USA}
\affiliation{Kavli Institute for Theoretical Physics, UCSB, Santa Barbara, CA 93106}

\begin{abstract}
We propose a mechanism to use nonlinearity arising from inter-particle interactions to significantly enhance rotation sensitivity of matter-wave interferometers. The method relies on modifying Sagnac interferometers by introducing a weak circular lattice potential that couples modes with opposite orbital angular momenta (OAM). The primary observable comprises of the modal population distributions measured at particular times. This provides an alternate mechanism for rotation sensing that requires substantially smaller ring size, even in the linear non-interacting regime. Nonlinearity can improve the sensitivity, as well as operation timescales, by several orders of magnitude.
\end{abstract}
\date{\today }
\pacs{ 03.75.Dg, 06.30.Gv, 67.85.De, 42.50.Tx}
\maketitle

{\em Introduction}: Atom interferometry has been pushing the limits of precision rotation sensing in recent years \cite{kasevich-2011,prentiss,kasevich-2006}. A common perception in the field is that the inherent atom-atom interaction, of the degenerate gases used as the working medium, is undesirable as being another source of systematic error. In this paper we show that the nonlinearity arising from such interactions can actually be utilized to significantly enhance the sensitivity of matter-wave rotation sensors. This work is motivated by a recent work on matter-wave transport \cite{das-PRA}, where nonlinearity was found to have a similar effect in linear geometry.

The primary operational principle in rotation sensors, atomic or optical, is the Sagnac effect. The {\em passive} version \cite{post-rmp} favored in atomic implementations measures the phase shift arising from the differential travel-time of counter-propagating waves in a rotating loop; while the {\em active} version, used in sensitive ring laser gyros \cite{chow-rmp}, utilizes the frequency splitting induced by rotation in oppositely travelling modes in a ring-shaped laser cavity.  Sagnac-based sensors, despite their long history, have seen scant considerations of positive utility of nonlinearity, with a few notable exceptions for optical realizations, such as a passive scheme proposal \cite{meystre} to use non-reciprocity of nonlinear susceptibilities for interfering beams of different amplitudes; and more recently there have been some advances \cite{shahriar,zhang} in increasing Sagnac sensitivity by manipulating the anomalous dispersion of light. But such considerations have not found extensions to matter waves.

Although it has been generally challenging to use the Kerr-type nonlinearity \cite{chow-rmp}, typical of both optical and cold atomic sensors, to enhance rotation sensitivity in the standard Sagnac configuration, we find a way to do so by introducing a crucial modification and choosing an alternate observable, distinct from the phase shift and the frequency splitting. There is, however, an indirect reliance on the frequency-splitting, in which it differs from current matter-wave rotation sensors that rely on phase shifts. Both attractive and repulsive non-linearities can be utilized.

\begin{figure}
\centering
\includegraphics[width=\columnwidth]{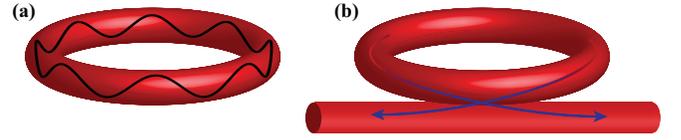}
\caption{(Color online) The atoms are trapped in a toroidal geometry with effectively one-dimensional dynamics about the major circle. (a) During operation a circular lattice potential exists as shown. (b) During readout, the lattice is turned off, and the counter-propagating modes are out-coupled to opposite directions of a tangentially coupled waveguide.}
\label{Figure-1}
\end{figure}

{\em Model and analysis}: We consider a Bose-Einstein condensates (BEC) in a toroidal trap as shown in Fig.~\ref{Figure-1}. Such a potential can be realized for example with orthogonal intersection of a Laguerre-Gaussian laser beam \cite{allen} with the waist of a Gaussian beam, both red-detuned with respect to the relevant atomic transition.  Such traps can sustain long-lived $\sim 40$ s condensate superfluid flow due to excitations of angular momentum eigenstates \cite{ramanathan}. Assuming the minor radius of the torus to be much smaller than the major radius $R$, as is possible with a higher order Laguerre-Gaussian beam \cite{wright,halkyard} and tightly focussed Gaussian beam, the dynamical description is taken to be one-dimensional (1D) along the major circle with excitations transverse to it frozen. The time evolution of the system with $N$ condensed atoms is then described by the 1D Gross-Pitaevskii equation in the interval $x \in [ 0,2\pi R) $, with periodic boundary conditions $\Psi(x=0,t)=\Psi(x=2\pi R,t)$,
\begin{eqnarray}
\label{Schr-equation}\hspace{-.3cm}
i\hbar\partial_{t}\Psi(x,t)=\left[-\frac{\hbar ^2\partial_{x}^2}{2m}+U(x,t)+g|\Psi(x,t)|^2 \right]\Psi(x,t),
\end{eqnarray}
where $m$ is the atomic mass and the nonlinear constant $g=Ng_{1D}$ is defined by the number of atoms {\it N}, and the effective 1D atom-atom interaction strength $g_{1D}$ \cite{das-PRL}.

For rotational sensing, the potential is taken to be of the form $U(x,t)=U(x-R\Omega t)$, which describes a stationary $U(x)$ rotating around the center of the circle with angular frequency $\Omega$. Transforming to polar coordinates in a rotating reference frame by setting $x\rightarrow y=x-R\Omega t$, $y\in[ 0,2\pi R)$, Eq.~\eqref{Schr-equation} becomes
\begin{eqnarray}
\label{rotating-Schr-equation}
&&i\hbar\partial_{t}\Psi(y,t)=\\
\nonumber
&&\left[-\frac{\hbar ^2\partial_{ y}^2}{2m}+i\hbar R\Omega\partial_{ y}+U( y)+g|\Psi( y,t)|^2 \right]\Psi( y,t),
\end{eqnarray}
with the new term $i\hbar R\Omega\partial_{ y}=-\Omega \hat{L}$, $\hat{L}$ being the angular momentum operator about the symmetry axis of the ring.

The Raman techniques of transferring quantized orbital angular momentum (OAM) from light beams to condensates \cite{wright08}, allow one to create circulating condensate with the wavefunction of the form $\left\langle y\ket{\pm l}\right.=1/\sqrt{2\pi R}\exp(\pm ily/R)$, $l\in\mathbb{N}$. When $U(y)\equiv 0$ and $\Omega= 0$, states $\ket{\pm l}$ are {\em degenerate} OAM eigenstates of Eq.~\eqref{rotating-Schr-equation}. This leads to the choice of the potential to be
\begin{eqnarray}
U(y)=U_0\cos^2(ly/R),
\label{potential}
\end{eqnarray}
a circular optical lattice \cite{padgett} (the circular  analog of optical Bragg gratings widely used to manipulate {\em linear} atomic momentum \cite{ketterle}). Such a potential can be created by interfering two counter-propagating Laguerre-Gaussian beams with opposite orbital angular momentum $\pm \hbar l$ per photon, possible in our case by partial retro-reflection of the trapping Laguerre-Gaussian beam.

{\em Two mode approximation}: We work in the limit of weak potential $U_0\ll \hbar^2l^2/2mR^2$, where the atom dynamics is effectively confined to a two dimensional subspace spanned by the states $\ket{\pm l}$, strongly coupled with each other, but negligibly coupled to other states.  Such a choice of optical potential results in the simplified atomic dynamics of an effective two-level system described by
\begin{figure}[h!]
\centering
\includegraphics[width=\columnwidth]{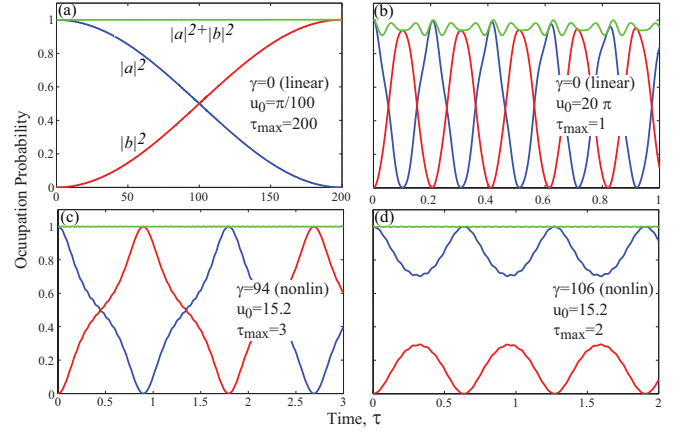}
\caption{(Color online) Time evolution of the populations $|a(\tau)| ^2$, ($|b(\tau)| ^2$) in states $\ket{\pm l}$ and their sum, is plotted by numerically solving Eq.~(\ref{rotating-Schr-equation}). The constancy of the sum justifies the two-mode model Eqs.~\eqref{two-level-rotating-Schr-equation-nonlin} $U_0\ll \hbar^2l^2/2mR^2$, assumed here; breakdown occurs  only at much higher lattice potentials (2000 times higher in (b)). With nonlinearity, there can be both (c) oscillatory and (d) self-trapped regimes. In all plots, $l=5$ and $R=10^{-5}$m.}
\label{Figure-2}
\end{figure}
\begin{eqnarray}
\ket{\Psi(t)}=a(t)\ket{l}+b(t)\ket{-l}.
\label{ansatz}
\end{eqnarray}
Inserting this into Eq.~\eqref{rotating-Schr-equation} and disregarding rapidly oscillating terms and global phase factors leads to a pair of coupled non-linear equations for the amplitudes:
\begin{eqnarray}
\label{two-level-rotating-Schr-equation-nonlin}
\nonumber
i\dot{a}(\tau)&=& \left[ -\omega l-\frac{\gamma}{2\pi}|a(\tau)|^2\right]  a(\tau)+ub(\tau),\\
i\dot{b}(\tau)&=& ua(\tau)+\left[ \omega l -\frac{\gamma}{2\pi}|b(\tau)|^2\right] b(\tau),
\end{eqnarray}
written in dimensionless form with redefined quantities:
\begin{eqnarray}
\nonumber
\tau\equiv t \hbar/(mR^2),&&\hspace{1cm}\omega\equiv\Omega mR^2/\hbar,\\
u\equiv u_0/4=U_0 mR^2/4\hbar^2,&& \hspace{1cm}\gamma\equiv gmR/\hbar^2.
\label{dimensionless-units}
\end{eqnarray}
The `dot' signifies derivative with respect to $\tau$. Our numerical results are obtained by solving the full Eq.~\eqref{rotating-Schr-equation} which confirm that the two mode-model is quite accurate, as indicated by the small deviation from the normalization $|a(\tau)|^2+|b(\tau)|^2=1$ seen in Fig.~\ref{Figure-2} for weak $U(y)$.

{\em Linear regime, $\gamma=0$}: When the atoms are non-interacting, which is possible by using Feshbach resonance to tune $\gamma\rightarrow 0$, Eqs.~\eqref{two-level-rotating-Schr-equation-nonlin} reduce to a pair of linear equations which can be exactly solved. Defining $\eta=\sqrt{u^2+\omega^2l^2}$, the solution  can be written as
\bn
\label{solution-two-level-rotating-Schr-equation}
a(\tau)=[ \cos(\eta\tau)+i\frac{\omega l}{\eta}\sin(\eta\tau)]a(0) -i\frac{u}{\eta}\sin(\eta\tau)b(0) \hspace{.3cm}\\
 b(\tau)= -i\frac{u}{\eta}\sin(\eta\tau) a(0)+  [ \cos(\eta\tau)-i\frac{\omega l}{\eta}\sin(\eta\tau)]b(0).\n
\en

\begin{figure}[t]
\centering
\includegraphics[width=\columnwidth]{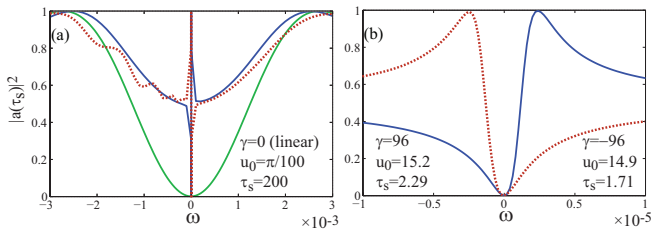}
\caption{(Color online) With $|a(0)|^2=1$, the population $|a(\tau_s)|^2$ of state $\ket{l}$ at a fixed evolution time $\tau_s$ depends on the angular velocity $\omega$ as shown. (a) Nonlinearity sharpens the dependence on the $\omega$ about a critical value, about $\omega=0$. (b) Details of the nonlinear profiles from (a) show that  the sensitivity to $\omega$ is over two orders of magnitude higher than for the linear case (note the $\omega$-axes scales). Time scales of operation are also shorter by a similar factor. In all plots, $l=5$ and  $R=10^{-5}$m; \emph{parameters shown in} (b) \emph{apply to nonlinear plots in} (a).}
\label{Figure-3}
\end{figure}

As seen from Eqs.~\eqref{two-level-rotating-Schr-equation-nonlin}, the lattice potential $U(y)$
leads to transitions between the states $\ket{l}$ and $\ket{-l}$, while the angular velocity $\omega$ lifts their degeneracy. We choose our initial state to be $a(0)=1,b(0)=0$. If there is no rotation, $\omega=0$, Eq.~\eqref{solution-two-level-rotating-Schr-equation} implies a complete population swap between the states at time $\tau_s=\pi/2u$, i.e.  $a(\tau_s,\omega=0)=0$, $b(\tau_s,\omega=0)=1$, as seen in Fig.~\ref{Figure-2}(a). But, for $\omega\neq 0$ the initial state remains partially populated:
\begin{eqnarray}
|a(\tau_s)|^2&=&\cos^2\left(\frac{\pi}{2}\frac{\eta}{u} \right)
+\frac{\omega^2l^2}{\eta^2} \sin^2\left(\frac{\pi}{2}\frac{\eta}{u} \right),
\label{probability}
\end{eqnarray}
due to the lifted degeneracy of states $\ket{\pm l}$ in the rotating frame.  The
dependence of $|a(\tau_s)|^2$ on the rate of rotation, $\omega$, is plotted in Fig.~\ref{Figure-3}(a) for the same parameters used in Fig.~\ref{Figure-2}(a). The clear dependence on $\omega$, indicates that the occupation probability after a fixed evolution time, $|a(\tau_s)|^2$ can used for  rotation sensing.

{\em Nonlinear regime, $\gamma\neq 0$}: The nonlinearity leads to an additional time-dependent potential due to the atomic density pattern which, in the two-mode picture, is $\propto [ab^*\exp(i2ly/R)+ c.c.]$. Since the nonlinear terms in Eqs.~\eqref{two-level-rotating-Schr-equation-nonlin} appear diagonally along with the rotational term, they directly influence the sensitivity to the rotation. General insight into the behavior can be deduced by noting that these equations  map to those for a non-rigid nonlinear pendulum \cite{wang}; thus, the dynamics displays both oscillatory and self trapped regimes as shown in Fig.~\ref{Figure-2}(c) and (d). Nonlinear strength above a critical point ($\gamma > 2\pi u_0$ for $\gamma>0$, $\omega=0$) causes self trapping that limits the transitions between states $\ket{\pm l}$ resulting in only partial transfer during a cycle; eventually there is no transfer at all for sufficiently high nonlinearity.

The nonlinearity can be used to significantly enhance rotation sensitivity by magnifying the dependence of the primary observable $|a(\tau_s)|^2$ on $\omega$ as we now explain. For a fixed potential $u$, and for $\omega=0$, the nonlinearity is tuned to be just below the critical point in the oscillatory regime and the time required for complete swap $\tau_s$ is set to be the detection time. Now, if there is rotation $\omega\neq 0$, there are two possibilities depending on the direction of rotation: (i) In one direction, the rotation tips system over into the self trapped-regime so that the $|a(\tau_s)|^2$ increases from $0$ for complete swap at $\omega=0$ to $1$ for no swap at all for a tiny change in $\omega$; (ii) When the rotation is in the opposite direction, the system stays in the oscillatory regime, but due to the proximity to the critical point, there is still a high sensitivity to small changes in $\omega$.  This behavior can be seen in Fig.~\ref{Figure-3}. Changing the sign of the nonlinearity switches the dependence on the direction of rotation in the relevant regime close to $\omega=0$, which follows from the relative signs of $\omega$ and $\gamma$ in  Eqs.~\eqref{two-level-rotating-Schr-equation-nonlin}.

The most remarkable feature apparent in the figures is that compared to the linear case, the sensitivity (for the assumed nonlinearity) is enhanced by over two orders of magnitude. An additional attractive feature implicit in the parameters shown in the figures, is that for experimentally feasible \cite{ramanathan} values of $\gamma$, the associated values of the lattice potential $u$ of interest are substantially higher than in the linear case; this has the effect of reducing the swap time $\tau_s$ by about two orders of magnitude as well. Thus, nonlinearity not only \emph{increases the sensitivity}, but the experiment can be accomplished in \emph{shorter times} as well.

\emph{Rotational Sensitivity}: The complementary probabilities of the two modes indicates a binomial (or Poisson) distribution, with each of the $N$ atoms constituting an independent trial. This can be brought about by tuning the nonlinearity to zero using Feshbach resonance during the detection process.
Therefore, a criterion for sensitivity to rotational frequency $\omega$ is provided by comparing the binomial mean value $ N|a(\tau_s)|^2$ and the standard deviation $\sqrt{N}|a(\tau_s)b(\tau_s)|$ for $\omega=0$, with that of the minimum value $\omega=\Delta\omega$ that is distinguishable from zero. This sets the criterion for frequency resolution and sensitivity
\begin{eqnarray}
\label{rotational-sensitivity}
N|a_0|^2+\sqrt{N}|a_0b_0|\leq N|a_{\Delta\omega}|^2-\sqrt{N}|a_{\Delta\omega}b_{\Delta\omega}|,
\end{eqnarray}
where $a_i$, $b_i$ are the respective amplitudes of the modes.

Using this criterion with parameters from the experiment in Ref.~\cite{ramanathan} with sodium atoms, $m=3.8\times 10^{-26}$ kg, $R= 10^{-5}$ m, $N=2\times 10^{4}$, for our choice of $l= 5$, we compute the short term sensitivity in the \emph{linear} case to be $\Delta\omega=1.1\times 10^{-5}$ which corresponds to $\Delta\Omega=2.9\times 10^{-4}$ rad/s, with a run-time $t_s=7.6$ s. In the nonlinear case, the experimental parameters lead to $\gamma=96$ as used here, and the sensitivity is significantly enhanced $\Delta\omega= 2.5\times 10^{-8}$ which corresponds to $\Delta\Omega= 6.5\times 10^{-7}$ rad/s, and the run-time shortened commensurately $t_s=0.087$s.  By repeating the cycles we can define a  process with cumulative atom-number and duration, so that on normalizing by the atoms used per second, $\Delta\Omega\times \sqrt{t_s}$,  we can frame our sensitivities in commonly used units: For the linear case $2.8\times 10^{-4}$ rad/s/$\sqrt{\rm Hz}$ and for nonlinear $1.9\times 10^{-7}$ rad/s/$\sqrt{\rm Hz}$.

\begin{figure}[h!]
\centering
\includegraphics[width=\columnwidth]{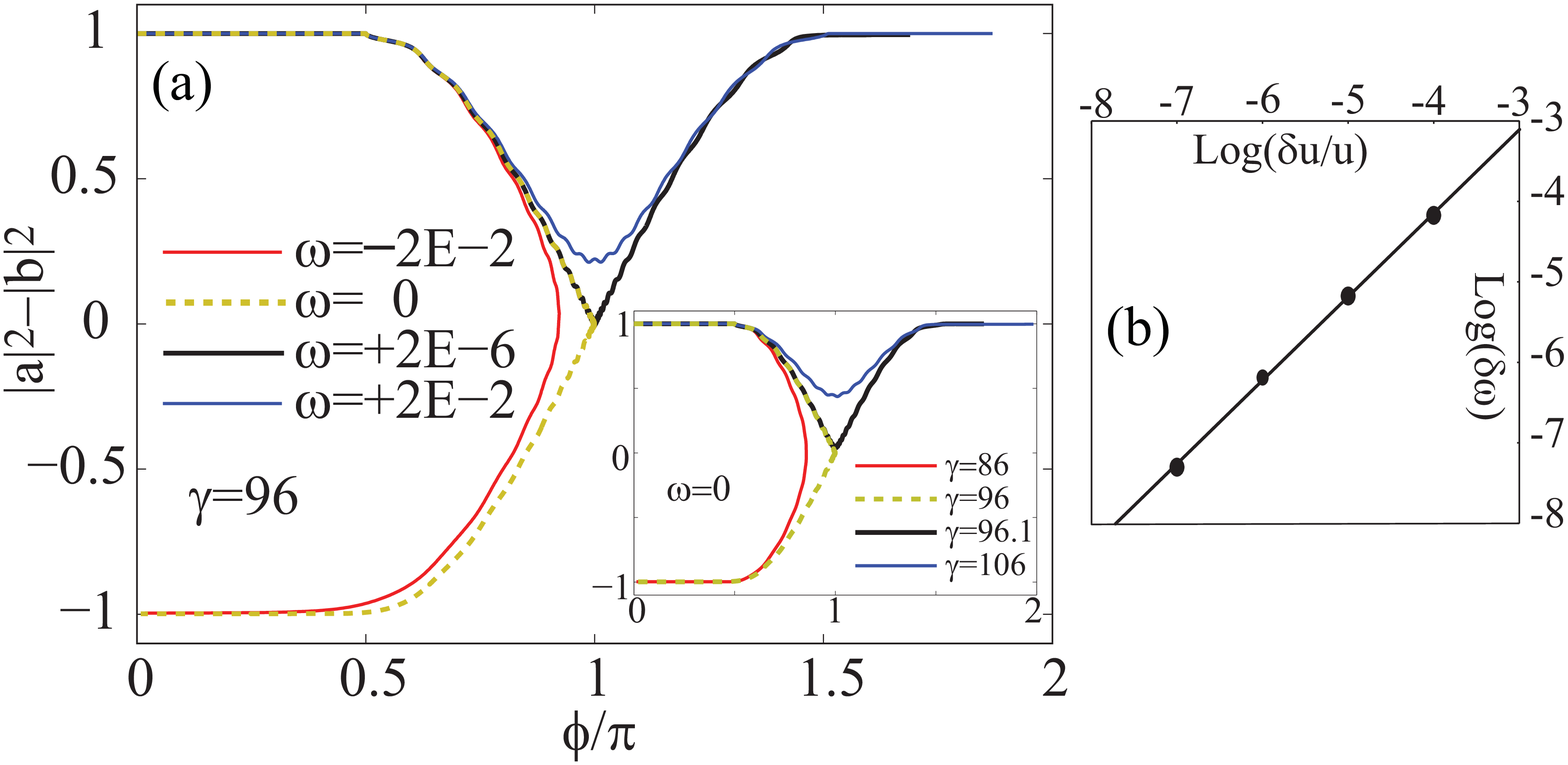}
\caption{(Color online)  Phase space diagram of probability difference versus phase difference of the mode amplitudes $a(\tau_s)$ and $b(\tau_s)$. Each line represents time evolution for specific values of $\gamma$ and $\omega$. (a) The inset shows transition from oscillatory to self-trapped behavior as interaction strength varies through the critical point at $\gamma\simeq 96.1$ for \emph{fixed} $\omega=0$. With the interaction strength \emph{fixed} at $\gamma=96$ instead (just below the critical point for $\omega=0$) in the primary plot, similar behavior occurs when $\omega$ is varied. (b) Deviation $\delta \omega$, of the location of the minimum in Fig.~\ref{Figure-3}, from $\omega=0$ is linearly proportional to the stability $\delta u/u$ of the lattice strength, the linear fit being $\log(\delta \omega)=1.04\times \log(\delta u/u)+0.011$.}
\label{Figure-4}
\end{figure}

In principle, by choosing the nonlinearity to be increasingly closer to the critical point, the sensitivity can be increased arbitrarily by causing more pronounced changes to $|a(\tau_s)|^2$ about $\omega=0$.  The limiting factor is the concurrent heightened sensitivity to the potential $u$. This means that any fluctuations in $u$ will change the position of the dip in Fig.~\ref{Figure-3}, so that for the same $\tau_s$ the minimum of the curves would be shifted $\delta\omega$ from $\omega=0$. In Fig.~\ref{Figure-4} we present a log-log plot of $\delta u/u$ versus $\delta\omega$, determined by numerical time-evolution for $u\pm\delta u$, and find that the dependence is essentially linear. In cold atom experiments stability of $(\delta u)/u\sim 10^{-6}$ can be reached. To access the enhanced sensitivity attainable with non-linearity, much better stability would be needed, such as attainable in some highly sensitive gravitational experiments, where stability of $(\delta u)/u\sim 10^{-8}$ is within reach \cite{rollins}.

{\em Comparison to the Sagnac effect:} If both the nonlinearity and lattice are absent, then  Eq.~\eqref{two-level-rotating-Schr-equation-nonlin} shows that rotation simply splits the energies of  $|\pm l\rangle$ levels by $2\omega l$, and we find that the corresponding frequency splitting in physical units is $\delta \nu=(4\Omega  A)/(\lambda P)$ which is \emph{identically the Sagnac frequency splitting} for matter waves, with $A=\pi R^2, P=2\pi R$ the area and perimeter of the ring and $\lambda=h/(mv)$ the de Broglie wavelength. This implies that the physical process in this limit is identical to the Sagnac effect.
Even without rotation $\omega=0$, the nonlinearity causes an energy splitting $(\gamma/2\pi)(|a(0)|^2-|b(0)|^2)$, (the populations remain constant due to lack of coupling in the absence of the lattice). This causes the standing wave formed by the counter-propagating modes to rotate.  Any physical rotation $\omega\neq 0$ also causes a similar rotation which will simply be added on to the nonlinearity-induced rotation. Since the nonlinearity itself does not depend on the rotation, it does not help in rotation sensing and is in fact undesirable because now the effect of the actual rotation has to be distinguished form a potentially much stronger fringe rotation due to the nonlinearity.

This makes it clear that the coupling introduced by the lattice is the crucial element in our proposed mechanism. It leads to the population oscillation that provides the alternative observable in the population of the modes, distinct from the phase shift or the frequency splitting that has been traditionally used in Sagnac-based rotation sensing. Crucially the coupling makes it possible for nonlinearity to have an enhancing effect on the sensitivity, which has not been possible in the traditional Sagnac setup.

{\em Detection of the motional states}: Rotational sensing in our scheme requires precise determination of the populations of states $\ket{\pm l}$. One possible way is indicated in Fig.~\ref{Figure-1}(b) that can work as follows: After duration $\tau_s$ of switching on, the lattice potential will be turned off and the nonlinearity $\gamma$ tuned to zero. This freezes the populations in the respective rotational states and avoids any complications due to nonlinearity during measurement. Then, a red-detuned Gaussian beam is coupled tangentially to the toroidal trap allowing the atoms to tunnel. The atoms in the two states $|\pm l\rangle$ will move out in opposite direction along the tangential beam and can be separately detected to yield the ratio of $|a(\tau)|^2$ and $|b(\tau)|^2$.

{\em Conclusions}: We have proposed a way to use nonlinearity to boost rotational sensitivity of matter-wave interferometers in a configuration where the observable comprises of populations of coupled modes at the end of a operation cycle of fixed time. This provides a way around the challenges faced in attempts to use nonlinearity to improve sensitivity of Sagnac interferometers. Even in the linear regime, our method offers high sensitivity levels that could be substantially improved with increasing particle number. Nonlinearity can boost the sensitivity to levels competitive with current matter-wave interferometers \cite{gauguet}, and notably this is achieved with a ring of area $\sim 10^{-4}$ mm$^2$ that is smaller by several orders of magnitude \cite{prentiss,kasevich-2011}. The nonlinear enhancement requires heightened stability of the mode-coupling potential, and can have applications impact in highly sensitive experiments where such stability is accessible.

{\em Acknowledgement}: M.K. and T.O. acknowledge the support of Czech Science Foundation Grant No. P205/10/1657. K.K.D. acknowledges valuable discussions with Y. Rostovtsev and support from the National Science Foundation under Grants No. PHY-0970012 and PHY11-25915.

\end{document}